\documentclass{iau} 
\usepackage{graphicx}

\title[Direct Imaging of Exoplanets at the Era of the Extremely Large Telescopes]
{Direct Imaging of Exoplanets at the Era of the Extremely Large Telescopes}

\author[Ga\"el Chauvin]
{Ga\"el Chauvin$^{1,2}$}

\affiliation{
$^{1}$ Unidad Mixta Internacional Franco-Chilena de Astronom\'{i}a, CNRS/INSU UMI 3386 and Departamento de Astronom\'{i}a, Universidad de Chile, Casilla 36-D, Santiago, Chile,\\[\affilskip] 
$^{2}$ Univ. Grenoble Alpes, CNRS, IPAG, F-38000 Grenoble, France.
\\email: {\tt gael.chauvin@univ-grenoble-alpes.fr}}

\pubyear{2015}
\volume{xxx}  
\jname{IAUS 347: Early Science with ELTs (EASE)}
\editors{A.C. Editor, B.D. Editor \& C.E. Editor, eds.}
\begin{document}

\maketitle

\begin{abstract}

Within ten years, the era of large-scale systematics surveys will decay thanks to a complete census of exoplanetary systems within 200\,pc from the Sun. With the first Lights foreseen between 2024 and 2028, the  new  generation  of  extremely large telescopes and planet imagers will arrive at a propitious time to exploit this manna of discoveries to characterize the formation, the evolution, and the physics of giant and telluric planets with the ultimate goal to search and discover bio-signatures. In that perspective, I will briefly summarize the main characteristics of the direct imaging instruments of the ELTs dedicated to the study of exoplanets, and I will review the key science cases (from the initial conditions of planetary formation, the architecture of planetary systems and the physics and atmospheres of giant and telluric planets) that they will address given their predicted performances. 
\end{abstract}

\firstsection 
\section{Introduction}

\begin{figure*}[t]
\begin{center}
 \includegraphics[width=\textwidth]{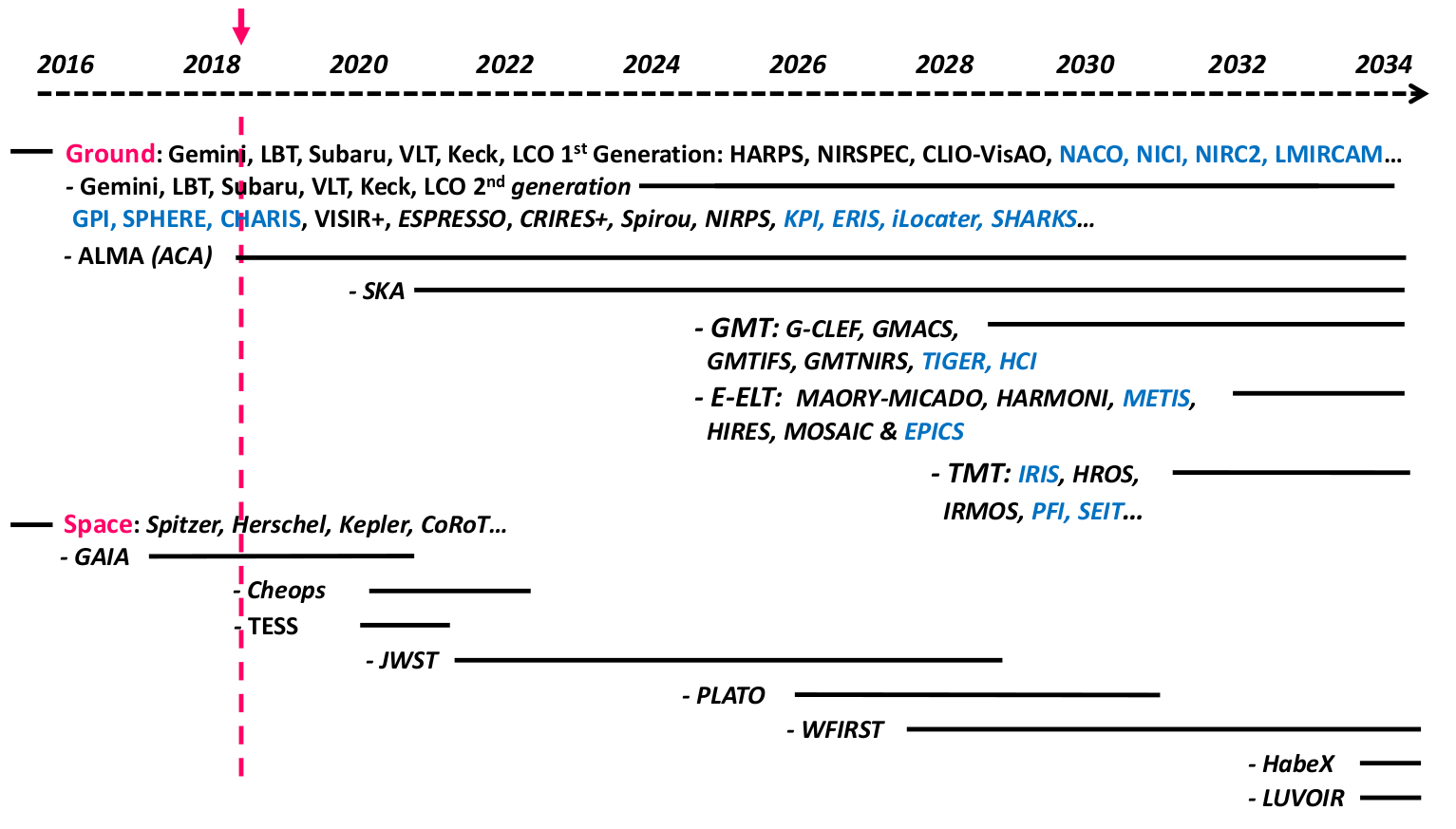} 
 \caption{Timeline of current \& future missions and instruments dedicated to exoplanets.}
   \label{fig:timeline}
\end{center}
\end{figure*}

The young field of exoplanetary science has exploded in the past
years.  Two decades ago, the only planets we knew were the ones of our Solar System. Today, thousands of exoplanets have been discovered since the 51 Peg discovery (\cite[Mayor \& Queloz 1995]{mayor1995}) and, with the diversity of systems found (Hot Jupiters, irradiated and  evaporating planets, misaligned planets with stellar spin, planets in binaries, telluric planets in habitable zone, discovery of Mars-size planet...), the theories of planetary formation and evolution have drastically evolved to digest these observing constraints.  Although on the timescale of a human Life, we may well be witnessing the first detection of bio-signatures in the atmosphere of a nearby exo-Earth at the horizon of 2030, we are still missing the full picture and some key fundamental questions lack answers regarding: i/ the existence of one or several mechanisms to form giant planets, ii/ the physics of accretion to form their gaseous atmospheres, iii/ the physical properties of young Jupiters and their time evolution, and iv/ the impact of dynamical evolution in crafting planetary system architectures. In that perspective, the upcoming decade is rich in terms of space missions
and ground-based instrumentation dedicated to this young field of modern
astronomy (see Fig.~\ref{fig:timeline}). 

With the new generation of planet imagers (SPHERE, GPI, SCExAO), the direct imaging technique now represents a unique path to characterize true analogs of cool Jovian planets orbiting at more than 5-10 au (\cite[Chauvin 2018]{chauvin2018}), a parameter space currently not explored by transit and radial velocity surveys. At sub-mm and centimetric wavelengths, ALMA in full capability now pursues the characterization of the cold dusty and gaseous component of young protoplanetary and debris disks with an exquisite spatial resolution (down to 0.1$~\!''$) and will be soon completed by the Square-Kilometer Array (SKA) starting in 2020. The arrival of a new generation of high-resolution spectrographs, ESPRESSO at VLT, CARMENES at CAO, SPIROU at CFHT now in operation, CRIRES+ at VLT, NIRPS at the ESO3.6m Telescope, and iLocater at LBT soon,  will extend the current NIRSPEC and HARPS horizon to the population of light telluric planets around solar and low-mass stars. In space, \textit{Gaia}, launched end-2013, will achieve a final astrometric precision of 10~$\mu$as in the context of a systematic survey of a billion of stars and therefore discover thousands of new planetary systems. The \textit{Gaia} data release 4 in 2022 should give us a complete census of the giant planet population between 2 and 4~au for stars closer than 200~pc. The new transiting
exoplanet survey satellite (\textit{TESS}, Transiting Exoplanet Survey Satellite) just started operation going beyond the \textit{Corot} and \textit{Kepler} missions with its first discovery of a transiting planet around $\pi$ Mensae (Huang et al. 2018). It is designed for a full-sky survey to reveal thousands of
transiting exoplanet candidates with the size of Earth or larger and
orbital periods of up to two months. This will be complemented by the
\textit{CHEOPS} (CHaracterising ExOPlanet Satellite) mission aimed at characterizing the structure of
exoplanets with typical sizes ranging from Neptune down to Earth
diameters orbiting bright stars (launch date in 2019). Finally,
\textit{PLATO} (Planetary Transits and Oscillations of Stars),
foreseen for 2024, will extend our knowledge on the content of
telluric planets at longer periods, up to several years, around relatively bright, nearby
stars. Within 10 years, the era of large-scale
systematics surveys will decay thanks to a complete census of
exoplanetary systems within 100--200~pc from the Sun. A new Era fully  dedicated to the characterization of known
systems will rise. Already initiated with \textit{Hubble}, \textit{Spitzer} and
the first generation of planet imagers and spectrographs, the
characterization of the physics of giant and telluric planets will intensify with
the operation of the James Webb Space Telescope (\textit{JWST}),
foreseen for the year 2021, which will address several key questions
for the study of young circumstellar disks and exoplanetary
atmospheres using direct imaging and transit and secondary eclipse
spectroscopy.

\begin{table}[!t]
\small
\caption{Instrumentation roadmap of the ELTs summarizing instruments and modes
adapted for high contrast imaging and spectroscopy of exoplanets and disks. For
AO flavors: seeing-limited (SL), single-conjugated AO with moderate-Strehl (SCAO) and extreme-AO
with high-Strehl performances (XAO). Various observing modes are proposed: imaging (IMG), medium
  and high resolution spectroscopy (MRS, HRS), integral field spectrograph
  (IFS), combined with coronography (corono) or polarimetry (polar).}
\label{tab:eelt}       
\begin{center}
\begin{tabular}{lllllllll}
\hline
Telescope                   & Instrument   & AO    & Mode   & $\lambda$  & Spectral     & FoV          & Add. \\
(\textit{1$^{st}$ Light})  &               &        &       & ($\mu$m)   & resolution   &  ($\,''$)  &           \\ 
\hline 
E-ELT                       & MICADO        & SCAO  & IMG    & $0.8-2.4$  & BB, NB       & $53$             & corono  \\
(\textit{2024})             &               & SCAO  & MRS    & $0.8-2.4$  & $<15000$   & $3-slit$               &              \\
                            & HARMONI       & SCAO  & IFS    & $0.5-2.5$  & 3500-20000   & $0.6$            & corono  \\
                            & METIS         & SCAO  & IMG    & $3-19$     & BB, NB       & $18$            & corono \\
                            &               & SCAO  & MRS    & $3-19$     & 5000         & $18$                &         \\ 
                            &               & SCAO  & IFS    & $3-13$     & 100000       & $0.4$            & corono \\
                            & HIRES        & SCAO  & IFS   & 0.33-2.4   & $<150000$    & $0.09$            &       \\
                            & EPICS         & XAO   & IFS    & 0.95-1.65  & 125-20000    & $0.8$          & corono \\
                            &               & XAO   & IMG    & 0.6-0.9    & BB, NB, DBI  & $2$          & polar \\
\hline
GMT                        & G-CLEF         & SCAO  & HRS    & $0.4-1.0$  & 20000-100000       & $7\times0.23$& corono  \\
(\textit{2024})            & GMTIFS         & SCAO  & IMG    & $1.0-2.5$  &   BB, NB           & $20$              &   \\
                           &                & SCAO  & IFU    & $1.0-2.5$  & 5000-10000      & $0.5\times0.25$   &   \\
                           & GMTNIRS        & SCAO  & HRS    & $1.1-2.5$  & 65000        & $1.2-slit$             &   \\
                           &                & SCAO  & HRS    & $3.0-5.0$  & 85000        & $1.2-slit$             &   \\
                           &  TIGER         & SCAO  & IMG    & $1.5-14.0$  & 300          & $30$              & corono  \\
                           &                & SCAO  & LRS    & $1.5-14.0$  & 300          & $30$              & corono  \\
\hline
TMT                        & IRIS           & SCAO  & IMG    & $0.85-2.4$  & BB, NB   & $33$              &   \\
(\textit{2028})            &                & SCAO  & IFS    & $0.85-2.4$  & 4000-8000   & $0.06\times0.5$   &   \\
                           & MICHI          & SCAO  & IMG    & $3-14$      & BB, NB      &   28                &  corono \\ 
                           &                & SCAO  & LRS    & $3-14$      & 600        &     $28-slit$              &  corono \\ 
                                                     &                & SCAO  & HRS    & $3-14$      & $<120000$        &     $2-slit$              &  corono \\ 
                           &                & SCAO  & IFS    & $7-14$   & 1000      &   $0.18\times0.07$               &  corono\\ 
                           & NIRES          & SCAO  & HRS    & $1.0-2.4$   & 20000-120000&     $2-slit$              &   \\ 
                           & PFI             & XAO   & IMG    & $1-2.5$   & BB, NB, DBI      &  $1$                 &  corono \\ 
                           &                & XAO   & IFS    & $1-2.5$   & 100          &    $1$               &  corono \\ 
\hline
\end{tabular}
\end{center}
\end{table}
\normalsize

\section{The era of ELTs}

Despite a reduced sensitivity compared to \textit{JWST}, the new
generation of extremely large telescope (ELT), the Giant Magellan Telescope
(GMT; \cite[Shectman \& Johns 2010]{shectman2010}), the Thirty Meters Telecope (TMT; \cite[Simard et al. 2010]{simard2010}) and the European Extremely Large Telescope (hereafter E-ELT; \cite[McPherson et al. 2012]{mcpherson2012}), will offer a unique spatial
resolution and instrumentation. With the first Lights foreseen between
2024 and 2028, these ELTs will arrive at a propitious time to exploit
discoveries of the upcoming generation of instruments and space
missions owing to its capabilities in terms of sensitivity, spatial
resolution and instrumental versatility. They will bring us a step
further in the understanding of the physics of the early phases of
stellar and planetary formation, the characterization of exoplanets,
and the quest of extraterrestrial Life. 


The three ELT projects (GMT, TMT and E-ELT) represent ones of the most challenging projects in
modern astronomy with the realization of 30 to 40m-class Adaptive Telescopes
designed for visible and infrared wavelengths and equipped with
segmented primary mirrors (7 segments of 8.4\, diameter  for GMT, 492 segments of 1.44\,m for TMT and 800 segments of 1.45\,m for E-ELT). They will cover a large variety of scientific topics
from the re-ionization of the early universe to the search for
bio-signatures on nearby exoplanets. They will therefore rely on a
diversity of dedicated instruments exploiting the telescope
unprecedented sensitivity and spatial resolution with various
observing modes: from integrated field spectroscopy, high-precision
astrometry, simultaneous multi-object spectroscopy of hundred of
targets, high resolution spectroscopy or high contrast imaging over a
broad range of wavelengths (from 0.33 to 19.0~$\mu$m) together with a
high operation efficiency. To respect these requirements, various
flavors of AO systems will be used to ultimately provide the international community with diffraction limited images down to $\sim10$~mas in $K$-band and/or to exploit the full patrol field of view of several arcminutes offered by these telescopes. These ELTs will undoubtedly open a new era for observers
to address various unanswered fundamental questions of astronomy: accelerating expansion of the universe, fundamental constants, galaxy formation and evolution, stellar populations, and the study of exoplanets and bio-signatures. The Table\,\ref{tab:eelt} summarizes the main instruments and modes that will address this ultimate science case combining high-angular resolution with adaptive techniques and/or imaging and low to high-resolution spectroscopy.


\section{Imaging and Characterizing Exoplanets with the ELTs}

\subsection{Initial conditions for planetary formation}

\begin{figure*}[t]
\begin{center}
 \includegraphics[height=5cm]{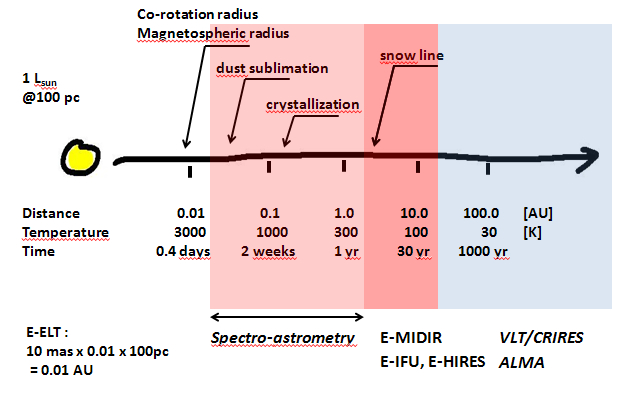} 
 \small
 \caption{Exploration space of the planet-forming regions by the ELTs}
 \normalsize
   \label{fig:inner}
\end{center}
\end{figure*}

The increased sensitivity and exquisite spatial resolution of the
ELTs will offer an incredible opportunity to scrutinize and
characterize the stellar environment from the inner regions of
proto-planetary disks, the chemistry of planet-forming zones, to the
characterization of recently formed giant planets in their birth
environment. Fig.~\ref{fig:inner} illustrates for instance the synergy between
current 10m-class telescope instrumentation, ALMA and the future instrumentation of the ELTs to characterize the spatial, temporal and chemical evolution of young proto-planetary disks. The combination of increased sensitivity, spatial resolution at the 10~mas scale and high spectral
resolution will open a new observing window at the sub-au scale to
study the physics of star-disk interactions.  With the use of
spectro-astrometric technique, a typical
astrometric precision of 100~$\mu$as will be achieved for a star at
100~pc, enabling a characterization of the physical processes of
accretion and ejection at a few solar radii ($1~R_\odot \sim
0.005$~au). This will place unique constraints on the star-disk
interactions processes, the role of magnetic fields (reconfiguration
and line reconnection) and the geometry of the accretion channels
close to the star. At such a scale, we will also explore the
properties of the inner circumstellar disks (asymmetries, warp,
puffed-up inner rim...). We will directly probe the Jet-launching and
stellar/disk winds regions (geometry and plasmas conditions), using
various spectroscopic emission and forbidden lines proxies of elements
like Hydrogen, He, Ca, Mg, Fe, O, N or S. Moreover, accessing these
close physical scales will enable short-term variability studies to
explore the temporal dynamics of physical processes over a few hours
and days timescale to witness in real time the evolution of the star-disk
interactions including accretion and ejection processes with the
magnetic field evolution.

\begin{figure*}[t]
\begin{center}
 \includegraphics[height=5cm]{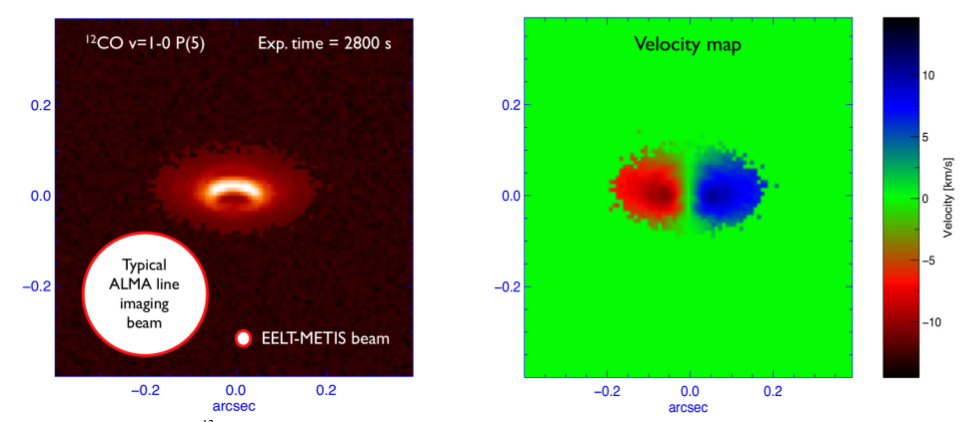} 
  \small
 \caption{\textit{Left}: Simulated $^{12}$CO line emission at 4.7\,$\mu$m of a
  proto-planetary disk, reconstructed with METIS. The map is
  continuum-subtracted and the velocity channels are optimally
  co-added. Also indicated is “typical” ALMA beam for line imaging of
  disks, estimated from Semenov et al. 2008. \textit{Right}: velocity map
  calculated as the first moment of the data cube. It shows that a
  resolving power of 100,000 (3 km.s$^{-1}$ is well matched to the spatial
  resolution of the E-ELT for a typical proto-planetary disk. Figure
  from Brandl et al. (2012).}
  \normalsize
   \label{fig:disk}
\end{center}
\end{figure*}

Direct spectral imaging of the planet-forming regions will be
achievable with the ELTs. For a typical young star at 100~pc, a
10~mas spatial resolution corresponds to physical separations of 1~au, i.e. to the exploration of the warm gas and dust spatial distribution down to the snow line, the disk evolution and dissipation to
ultimately determine the initial conditions of planetary
formation. Fig.~\ref{fig:disk} presents simulations of the young
proto-planetary disk SR\,21 (Ophiucus, 160pc, 1 Myr) seen in spectral
imaging in the $^{12}$CO(1-0) line at 4.7~$\mu$m with the IFU mode of METIS (\cite[Brandl et al. 2012]{brandl2012}). The inner CO gaseous gap at 18~au is
directly resolved.  In addition to directly image the gas
distribution, asymmetries and over densities in planet-forming zones,
the spectral information with a resolving power of 100,000 will enable
to directly map the gas dynamics with a velocity precision of
3~km.s$^{-1}$. Keplerian rotation will be distinguished from
wind, accretion or Jet components. Deviation from Keplerian might also help to distinguish the presence of hidden planets as recently done with ALMA for HD\,163296. The ELTs will therefore combine high angular resolution and high spectral resolution with instruments like METIS and HIRES at the E-ELT, G-CLEF and GMTNIRS at the GMT or HROS, MIRES and NIRES on the TMT to uniquely explore the inner planetary-forming regions ($\le20$~au). They will optimally complement ALMA or SKA observations,
more sensitive to the characterization of the cold, outer dust and gas components of circumstellar environments ($20-100$~au).

In addition to the disk spatial structure and kinematics, these instruments will directly explore the inner disk chemistry, the disk atmospheres, the physical transport of volatile ices either vertically or radially, and the
importance of non-thermal excitation processes in the planet-forming regions. They will offer a direct view of the
distribution and the dynamics of water, playing a key role for the
planetesimals formation and the disk cooling. It will also give clues on the organics content like CH$_4$, C$_2$H$_2$, HCN in the planet-forming regions and the prebiotic chemistry. The study of isotopic fractionation should also enable to probe the chemical and physical conditions in the proto-planetary disks to improve our understanding of the transfer processes of water on telluric planets, including therefore the Earth itself. It represents of course a mandatory step to understand the formation of favorable conditions for life on telluric planets.


\begin{figure*}[t]
\begin{center}
 \includegraphics[height=5cm]{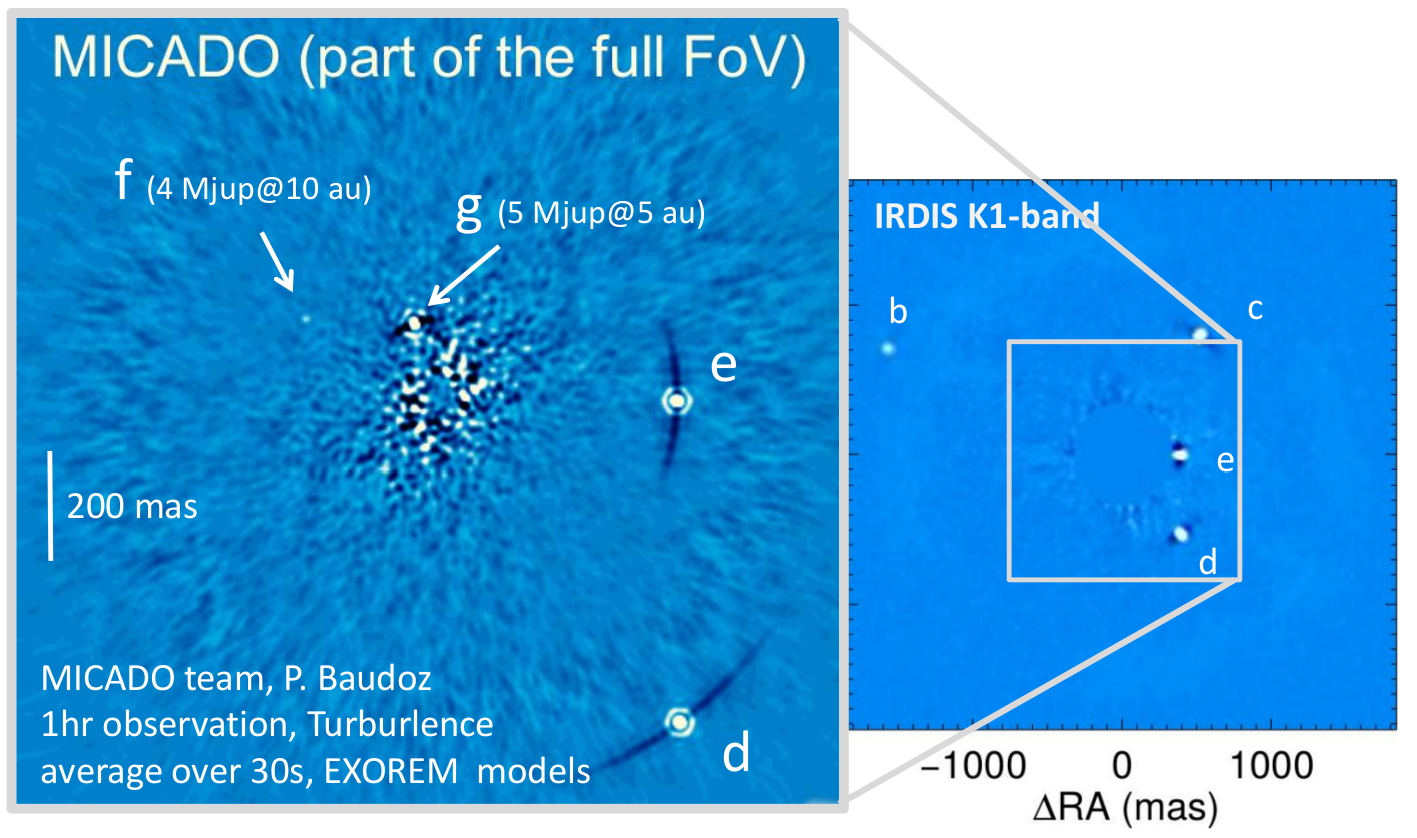} 
 \includegraphics[height=5cm]{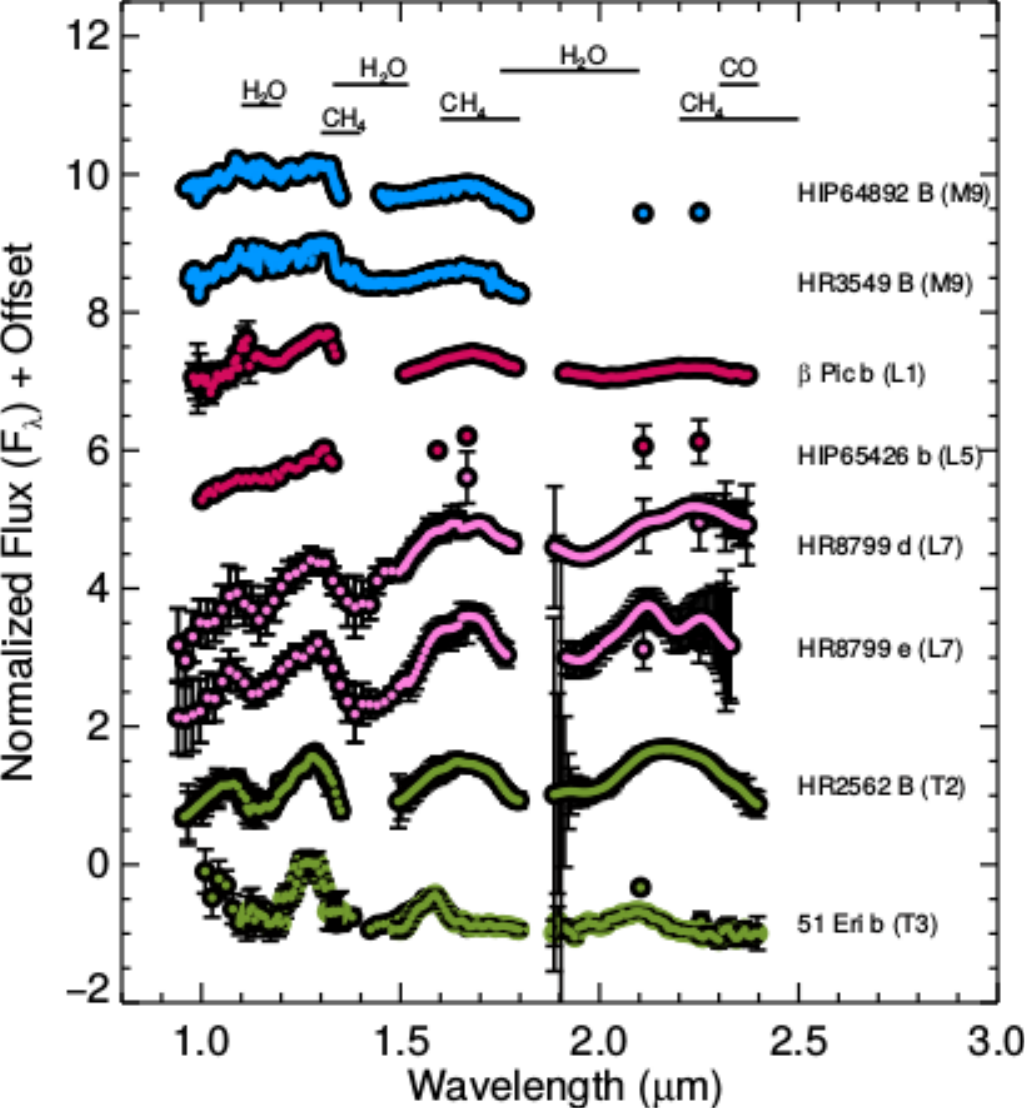} 
  \small
 \caption{\textit{Left:} MICADO simulations of Direct Imaging observations of the HR\,8799 planetary systems leading to the detection of putative inner planets $f$ and $g$ to illustrate the contrast performances already offered with the first Light instruments of ELTs (Priv. comm. P. Baudoz, MICDO team). \textit{Right:} Sequence of exoplanet's spectra characterized in the course of the SHINE and GPIES surveys (Priv. comm GPIES and SHINE teams). }
\normalsize
   \label{fig:synergy}
\end{center}
\end{figure*}

\subsection{Architecture of planetary systems}

With the first Light instruments, the combination of high angular resolution and moderate high contrast ($10^{-6}$ at $\sim200$\,mas) will enable the detection of fine gaseous and dusty structures in proto-planetary and allows to directly detect giant planets recently formed in their birth environment. This will be particularly crucial to understand the physics of gas accretion determining the ultimate physical properties of giant planets, but also the physics of planet-disk interactions. It is now clear that planetary formation has a connection with the various spatial structures and asymmetries (warp, cavity, spiral, hole, vortex...) observed in young disks. Instruments like METIS, HIRES, EPICS at E-ELT or MICHI, PFI at TMT and TIGER at GMT will ideally probed with high-contrast imaging (contrast goals ranging from $10^{-6}$ at 200~mas to $10^{-9}$ at 20~mas), from visible to mid-IR, the existence of young giant (and potentially telluric) proto-planets down to the snow line around young stars together with the disk structures and physical properties (see Fig.\,4, \textit{Left}). At the more evolved stage of debris disk, the direct imaging of young solar system analogs, i.e. multi-belt architectures hosting imaged planets like HR\,8799\,bcde, $\beta$ Pictoris b or more recently HD\,95086\,b, will be precious to understand the origin of our own solar system and test if giant planets might play a crucial dynamical role in the formation of telluric planets in stable habitable zones.

Ultimately, the synergy offered by the instrumentation of ELTs with other transit, radial velocity, astrometric and $\mu$-lensing surveys will enable a global exploration of the planetary system architecture at all orbits (for the giant down to the telluric population; see Fig.\,5). The complete overlap of these techniques will enable a systematical characterization of their frequency, multiplicity, distribution of mass and orbital parameters (period, eccentricity) for a broad range of stellar properties (mass, metallicity and age). Observables will be directly confronted to predictions of population synthesis models for various types of formation mechanisms (core accretion, gravitational instability or gravo-turbulent fragmentation). This will allow identifying the key mechanisms of formation and dynamical evolution (planet-disk and planet-planet interactions) of planetary architectures. This is an essential step toward the understanding of the material redistribution in young planetary systems. This is directly connected with the transport of planetesimals from beyond the ice-line bringing water and organic molecules to the inner telluric planets (as proposed for Earth with the Late Heavy Bombardment of the young solar system).

\begin{figure*}[t]
\begin{center}
\hspace{-1.4cm}
 \includegraphics[height=7cm]{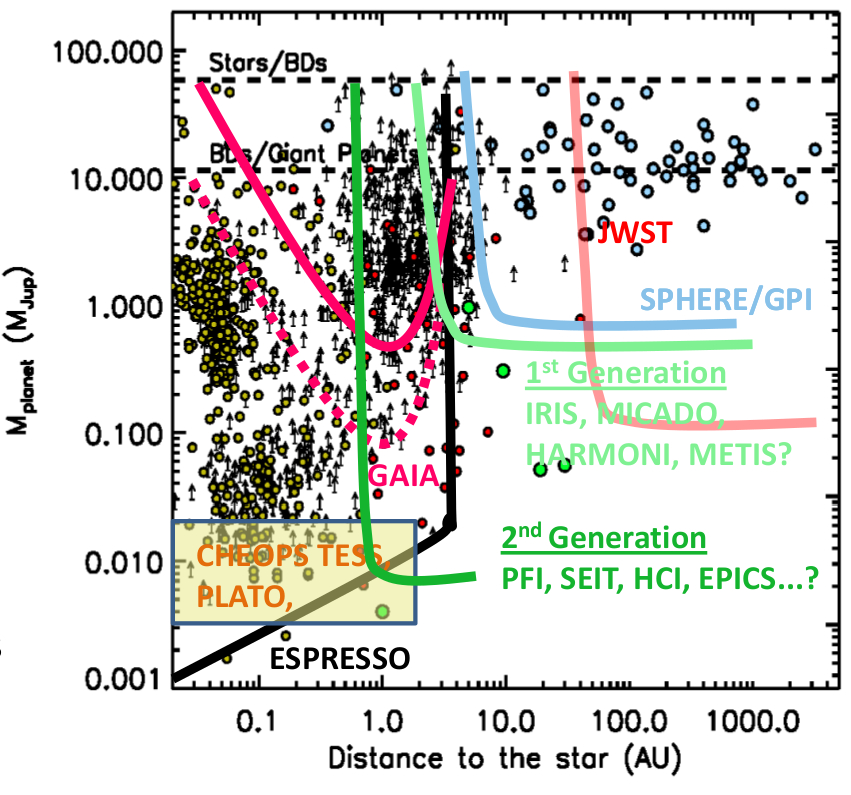} 
  \small
 \caption{Confirmed exoplanets discovered today with the indicative detection limits of
forthcoming planet hunting instruments and space mis-
sions together with the first and second generation of instruments of the ELT.}
\normalsize
   \label{fig:synergy}
\end{center}
\end{figure*}

\subsection{Physics of exoplanets}

Finally, first Lights of ELTs will arrive at a propitious time where thousand of new planetary systems will have been discovered and characterized by a large set of instruments or space missions, therefore covering a broad
range of the parameter space in terms of physical properties
(planetary masses, semi-major axis, radii, density, luminosity,
atmospheric composition) and stellar host properties (age, mass,
binarity, composition...). ELTs will therefore be mainly used for the fine characterization of known planetary systems and benefit from enhanced sensitivity, spatial resolution and instrumental versatility. The fact that observing techniques like direct imaging and radial velocity will overlap for the first time in the planetary mass regime will enable the simultaneous determination
of the planet's mass and luminosity (see Fig.~\ref{fig:synergy}). It
will therefore set fundamental constraints on the gas accretion
history of giant planets, therefore their mechanisms of formation and
evolution. It will also enable to characterize the atmosphere of giant
exoplanets and potentially be sensitive enough to probe super-Earths
and Exo-Earths.

The era of the characterization of exoplanets has already started a
decade ago with the atmospheric characterization of hot and strongly
irradiated Hot Jupiters like HD\,209458 (\cite[Charbonneau et al. 2012]{charbonneau2002}). Such observations have been reported now for over 30 exoplanets to date, including hot Jupiters, hot Neptunes, and even super-Earths. The presence of water, carbon monoxide and methane
molecules, of haze revealed by Rayleigh scattering, observation of
day-night temperature gradients, constraints on vertical atmospheric
structure and atmospheric escape have been evidenced in the past
decade. More recently, VLT observations with CRIRES
at high-spectral resolution (hereafter refered as high-dispersed
spectroscopy, HDS) showed that spectral features from planetary
atmospheres can be disentangled from telluric and stellar lines making
use of the radial velocity variations of the exoplanet. Planet to star contrast of $10^{-4}$ to $10^{-5}$
are typically expected for Hot Jupiters around solar-type stars. Such
a contrast would be also sufficient to explore the atmosphere of a
Earth-twin orbiting a small red dwarf (albeit such a star will be orders of magnitude fainter which imply
decade of transit observations). In addition to classical transit and
secondary eclipse low-resolution spectroscopy of exoplanets down to
telluric masses, the use of HDS with instruments like HIRES, METIS, G-CLEF, GMTNIRS, MICHI or NIRES will
enable to directly resolve the molecular lines (in transmission,
emission but also reflection) of giant planets and to search for the
signatures of CO, CO$_2$, H$_2$O, CH$_4$ and possibly NH$_3$, directly map their spatial distribution using Doppler imaging techniques (see Fig.\,6, \textit{Left}; \cite[Crossfield et al. 2014]{crossfiled2014}), and even
constrain the planet's albedo. One might expect to get constraints on
the structures and dynamics to characterize the processes of
inversion, vertical mixing, circulation and evaporation of the
planetary atmospheres of giant and telluric planets.

\begin{figure*}[t]
\begin{center}
 \includegraphics[height=5cm]{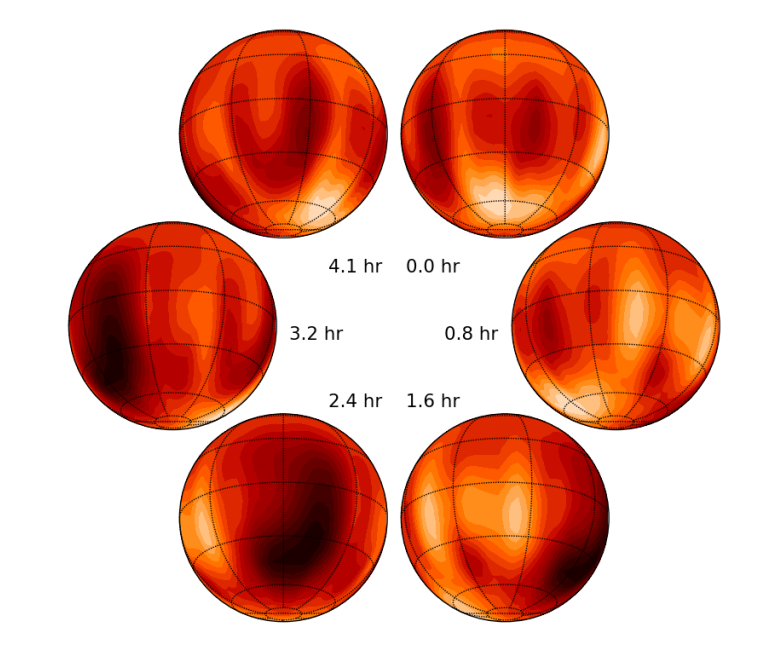} 
 \includegraphics[height=5cm]{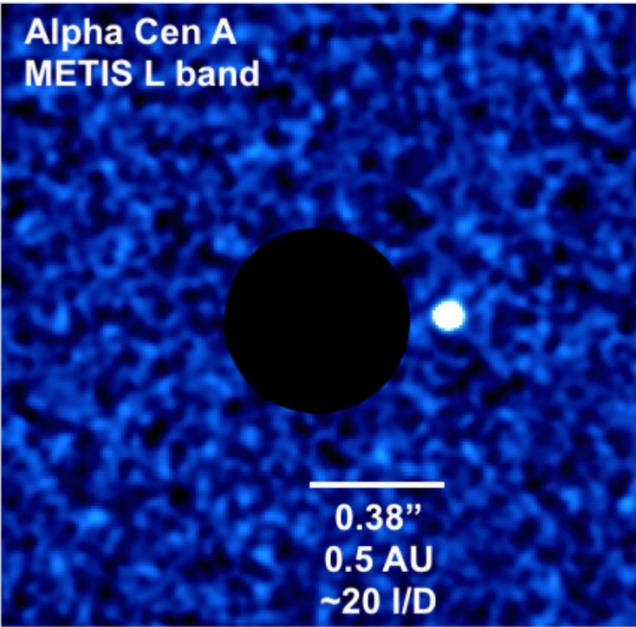} 
  \small
 \caption{\textit{Left,} a global cloud map of the nearest known brown dwarf Luhman 16 B. Figure from \cite{crossfield2014}. \textit{Right,} HDS + XAO cross-correlation map of 30 hours of optical obser-
vations with METIS at the E-ELT of Alpha Cen A, 1.3\,pc. An Earth twin, CO, CH$_4$ \& H$_2$O-rich, is detected at 0.5 au. Figure adapted from \cite{snellen2015}.}
\normalsize
   \label{fig:synergy}
\end{center}
\end{figure*}

At longer periods, high-contrast low to medium-resolution spectroscopy in direct imaging
already enabled to initiate the characterization of cool giant planets (see Fig.\,4, \textit{Right}). Exploiting that same
technique, MICADO, HARMONI, METIS, TIGER, PFI, EPICS will offer a
complementary view to directly resolve spatially and spectrally the
photons of cool giant exoplanets from massive Super-Jupiter down to
exo-Saturns and possibly Super-Earths. The study of their atmosphere
composition and element abundances relative to the stellar ones will
be precious to constrain the formation mechanisms of giant planets for
instance. The
ultimate goal will be to reach the necessary contrast of about $10^{-9}$ at angular separations of 20~mas around nearby M
dwarfs (see Fig.\,6, \textit{Right}). It represents the very ambitious specification to meet to detect and
characterize Super-Earths and Exo-Earths in the Habitable Zone
(0.02~au at 10~pc around M dwarfs). Pushing this limit with ELTs
is directly linked with the motivation to detect bio-clues like
O$_2$, O$_3$, CH$_4$, CO$_2$ and H$_2$O in the atmospheres of telluric planets in habitable zones. A very promising approach arises from the synergy of XAO and HDS techniques \cite[Snellen et al. 2015]{snellen2015}). It
represents a very exciting perspective in terms of future development in the view of future planet direct imagers like METIS, MICHI, PFI, EPICS that could lead to the
first characterization of exo-earths and possibly the first probable discovery
of exo-life, a scientific breakthrough of great philosophical value.

\begin{discussion}

\discuss{Lynne Hillenbrand}{At wide separations, there is a rather large gap between the nominal sensitivity of the operating instruments SPHERE and GPI - Each of which have observed hundreds of stars - and the detect giant planets. Specifically, the gap appears between 0.7 and 7\,M$_{Jup}$ at separation larger than /\,au. Do you have any comment? Can it be concluded that the objects in this mass and separation range do not exist? Or is there perhaps some over-estimation of the current instrument sensitivity?}
\discuss{Gael Chauvin}{The SPHERE and GPI sensitivities presented in this talk (and in Fig. 4 of this proceeding) are rough estimation of the current mass and separation parameter space probed by current XAO imagers for comparison with other techniques. The goals are really to illustrate the complementary of the different techniques and to highlight the typical gains expected in separation and contrast with the first Light instruments and the 2$^{nd}$ generation of the ELTs. They actually aimed at reaching the ambitious contrast goals of $10^{-6}$ at 200~mas down to $10^{-9}$ at 20~mas. More accurate statistical determinations of the detection probabilities achieved by the SPHERE and GPI surveys are currently underway. They should soon provide us with the detectio probability space explored, and, considering the number of discoveries, the frequency of giant planets found in the mass and separation domain. Results
of the first generation of planet imagers (NaCo, NIRC2, NICI, HiCIAO...) on 10-m class telescopes were mostly sensitive to giant planets more massive than $5\,M_{\rm{Jup}}$ for semi-major axis between typically 30 to 300\,au.
Pushing that logic, the recent large meta-analysis of 
384 unique and single young (5--300 Myr) stars spanning stellar masses between 0.1 and $3.0\,M_{\odot}$ and combining several DI surveys (\cite[Bowler 2016]{bowler2016}) illustrates that the corresponding overall occurrence rate of $5-13M_{\rm{Jup}}$ companions at orbital distances of 30--300 au remains relatively low with frequencies of planets orbiting BA, FGK, and M stars of ${2.8}_{-2.3}^{+3.7}\%$, $<4.1\%$, and $<3.9\%$, respectively.}

\end{discussion}

\end{document}